# Enhancing Ionic Conductivity of ceramic $Na_3Zr_2Si_2PO_{12}$ via NaI Substitution

Aditya Sekhar[a], Nikesh Lilani[a], and M. Dinachandra Singh[b,c],

[a]*Department of Physics, NIT Rourkela*
[b]*Student Research Exposure Lab (SUREELA), Shiksha Sopan*
[c]*Department of Sustainable Energy Engineering, Indian Institute of Technology Kanpur, Kanpur 208016, India*

Corresponding Author Email: m.dinamangang@gmail.com

**Abstract:**

The high ionic conducting sodium superionic conductor $Na_3Zr_2Si_2PO_{12}$ -NaI composites have been prepared successfully via solid state reaction route. As the NaI content increases, the ionic conductivity significantly changes. The composites with 4% NaI show a maximum ionic conductivity of ~ 3 x $10^{-4}$ $\Omega^{-1}cm^{-1}$ at 200°C which is one order of magnitude rise as compared to pristine sample. Further crystal structure and surface morphology also reveal the formation of a glassy phase due to the presence of Na-Si-P-O-I interaction.

1. **Introduction:**

Electrolytes play a significant role in determining the battery parameters like energy density, life span, etc., of the battery. Both solid and liquid electrolytes could be used. Currently, liquid electrolytes are prominently used due to the ease of operation and low cost, but solid electrolytes are good alternatives because of their non-flammability, non-volatility, and greater thermal stability [1–3]. Also, solid electrolytes have a greater electrochemical potential window than liquid state electrolytes. Due to these advantages of solid-state electrolytes, there has been a lot of focus on developing All Solid-state batteries (ASSB) [4]. For developing ASSB, it's crucial to find a solid electrolyte that has an ionic conductivity comparable to the ionic conductivity of widely used liquid electrolytes and also has stability with electrodes during charging and discharging. NASICON electrolyte is one of the leading contenders for solid electrolytes. NASICON electrolyte, with stoichiometry $Na_{1+x}Zr_2Si_xP_{3-x}O_{12}$ (0=<x<=3) (NZSP) is a very well-known and widely studied electrolyte material [5,6]. It has high $Na^+$ conductivity and a wide electrochemical potential stability window, thus making it a suitable choice for the solid-state electrolyte [7,8]. NZSP forms a 3D ionic conduction path that consists of $SiO_4$ and $PO_4$ tetrahedra that share corners with octahedral $ZrO_6$ [9,10]. It has a rhombohedral phase for 0 <=x<=3 except for 1.8 <=x<= 2.2 where the monoclinic phase is stable at room temperature[11].

There has been a lot of work done in enhancing the ionic conductivity of NZSP. Han Wang et. al. [12] studied the synthesis of NZSP with $Na_2SiO_3$ additives by liquid phase sintering and reported that NZSP with 5 wt% of $Na_2SiO_3$ achieved the highest ionic conductivity of 1.28 x $10^{-3}$ $Scm^{-1}$. Shengnan He et. al.[13] studied the F- assisted NZSP electrolyte and reported a high conductivity of 1.41x $10^{-3}$ $Scm^{-1}$ at 27°C as compared to unassisted NZSP. Eunseok Heo et. al. [14] studied the effects of potassium substitution and reported an ionic conductivity of 7.7 x$10^{-4}$ $Scm^{-1}$, 2 times greater than that of the undoped sample.

The present study focuses on enhancing the ionic conductivity of NZSP. We added NaI to the NZSP mixture in varying weight percentages (4%,6%,8%, and 10%), and found that the mixtures with 4% and 6% NaI showed greater ionic conductivity than pure NZSP. In this paper, we discuss the preparation and characterization techniques for the same.

**2. Synthesis and characterization:**

NASICON structured $Na_3Zr_2Si_2PO_{12}$ (NZSP) were prepared by conventional solid-state reaction route [15,16]. In this process, all the constituents' compounds of NASICON viz. $Na_2CO_3$, $ZrOCl_2$, $SiO2$, $NH4H2PO4$, and Sodium iodide were taken in a stoichiometric weight ratio and further grounded into a fine homogenous powder using an agate mortar and pestle for ~ 15 min. The obtained mixture was transferred into an alumina crucible, and calcined at 700°C for 2 hrs and followed by sintering at 1000°C for another 2 hrs. Similarly, $Na_3Zr_2Si_2PO_{12}$ complexed with ionic salt (NaI) in a configuration of NZSP-xNaI, where x corresponds to 0-12 wt% was also prepared. Further to study the electrical properties of the compound, the obtained sample was again grounded into fine particles ~ 5min, followed by pelletization at 1000 PSI pressure for 10 min using a KBr pelletizer, and stainless steel die of 10 mm diameter. To release the residual stress due to pressurization, the pellets were annealed at 600°C for 6 hr. The obtained pellet was kept in a vacuum desiccator for further characterization. The ionic conductivity of the composites was studied using a scientific LCR meter (4Hz-25 KHz,) at a temperature range from 50°C to 300°C (in-house built sample holder with the furnace). Further crystal structure and surface morphology of the composites were studied using X-Ray diffraction and filed emission scanning electron microscopy technique, respectively.

## 3. Result and Discussion:

X-ray diffraction (XRD) measurements were carried out for the composites of NaI-NZSP and, are shown in Fig. 1. As observed in fig., the crystalline peaks at $20^0$, $23^0$, $27^0$, and $31^0$ confirm the formation of NZSP monoclinic structure. However, it is also observed some impurities correspond to $Na_2ZrSiO_5$ and $SiO_2$. This impurity might be due to insufficient sintering temperature and timing. As evident in Fig., the XRD crystalline peaks seems to be suppressed with NaI content. Such glassy phase formation is may due to the presence of Na-Si-P-O-I interaction. However, the XRD pattern does not show any significant $I^-$ dominant related peaks. Further to understand the effect of NaI, FESEM images were taken and shown in Fig.2.

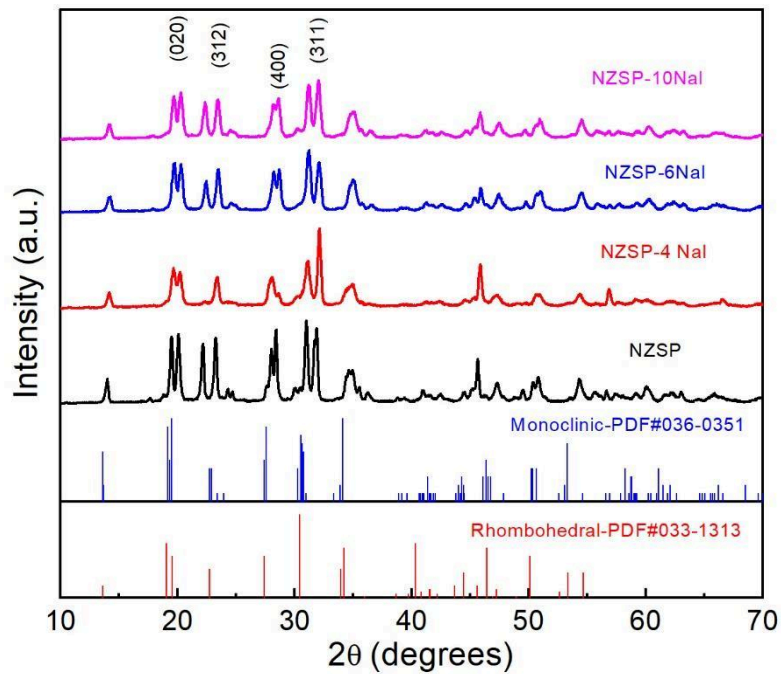

***Fig 1***: *XRD patterns of NZSP+ x% NaI (x=0, 4, 6, 10) samples.*

Fig.2 and b show FESEM images for pristine NZSP and 4% NaI content composites respectively. As evident in Fig.2, both composites show the homogenous distribution of particles having a dominantly cubic structure. Also, the sample with 4% NaI content exhibited glassy behavior as compared to pristine NZSP. This further compliments the above findings in XRD. However, to study the effect of such a structure in ionic conducting, temperature-dependent ionic conductivity measurements were carried out.

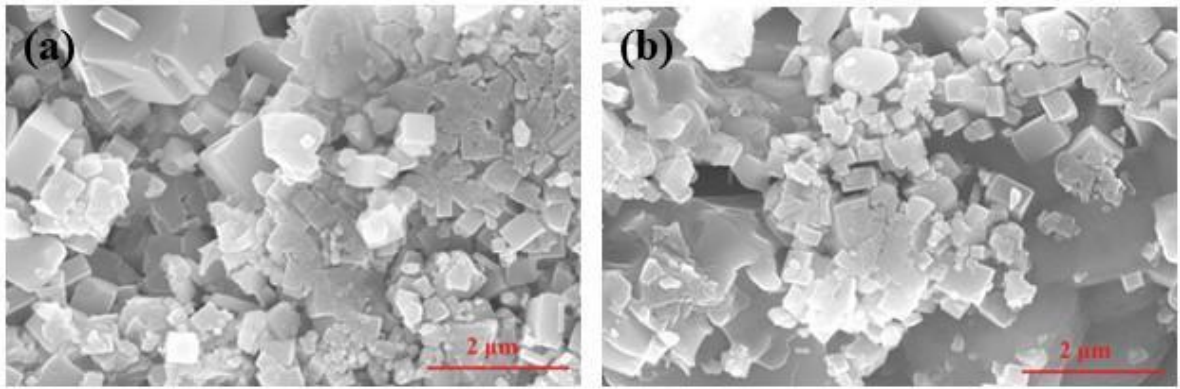

**Fig.2** FESEM images of (a) Pristine NZSP and (b) 4% NaI content NZSP composites

Fig. 3a shows temperature-dependent Nyquist plot obtained in a frequency range of 100Hz to 25kHz for pristine NZSP. In general, NZSP composites exhibited a depressed semi circle followed by a spike at a lower frequency. However, due to insufficient frequency range, the Nyquist plot shows an incomplete depressed semicircle as shown in Fig 3a. Through electric circuit component fitting (inset of Fig. 3a), the corresponding resistance was studied and fitted data is shown in Fig. by line. The resistance R1 and R2 can be assigned as contact resistance and overall bulk resistance of the sample (i.e. sum of Grain boundaries resistance and in-grain resistance).

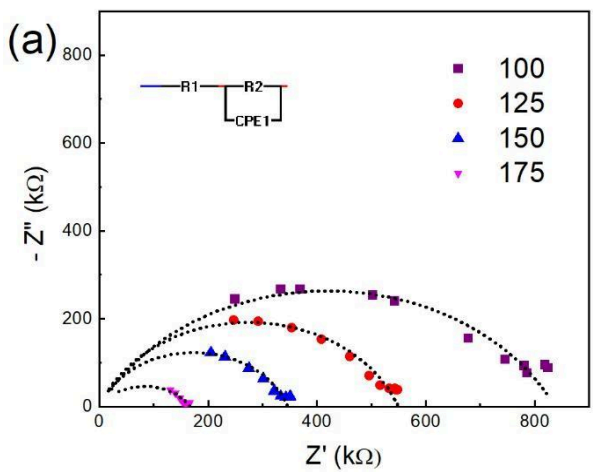

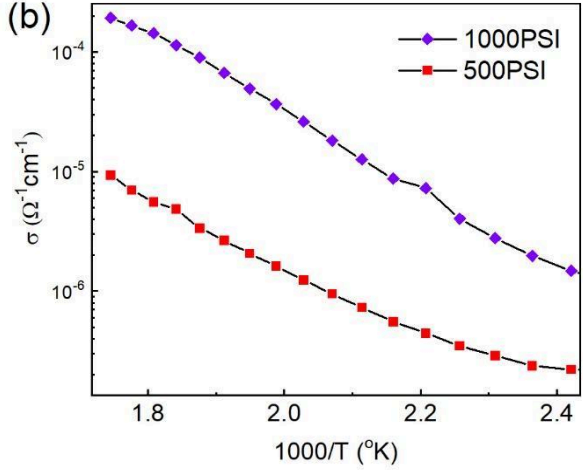

***Fig-3:*** *(a) Temperature-dependent Nyquist plot for pristine NZSP sample and (b) Temperature-dependent Ionic Conductivity of NZSP samples prepared at 500 PSI and 1000 PSI respectively.*

Thus in further studies, impedance was measured at 1 KHz where the dc conductivity region lies. To study the importance of pelletization pressure on ionic conductivity, pressure-dependent ionic conductivity was studied and plotted in Fig. 3b. As evident in Fig. 3b, pristine NZSP sample prepared at 1000 PSI, shows higher ionic conductivity with one order of magnitude rise as compared to the sample prepared at 500 PSI. This is due to the greater densification of NZSP particles.

The effect of NaI content on ionic conductivity was studied and the temperature-dependent ionic conductivity with NaI content is shown in Fig. 4a. As evident in fig., the temperature-dependent ionic conductivity exhibited Arrhenius behavior for all the composites. And thus suggest the ion's transport mechanism is due to hopping. It is evident that the ionic conductivity gradually increases to 8% NaI content. The composites with 4% NaI content shows the highest ionic conductivity of ~ 3 x $10^{-4}$ $\Omega^{-1}cm^{-1}$ at 200°C which is one order of magnitude rise as compared to pristine NZSP. The rise in the ionic conductivity is may be due to the glassy phase introduced by F$^-$ ion as described in XRD and FESEM. This glassy phase might provide pathways for the ion to conduct. Also, the activation energy decreases with NaI content as shown in Fig. 4b. The results complement the rise in ion ionic conductivity due to glassy phase formation as its amorphous nature provides a smooth ion hoping process.

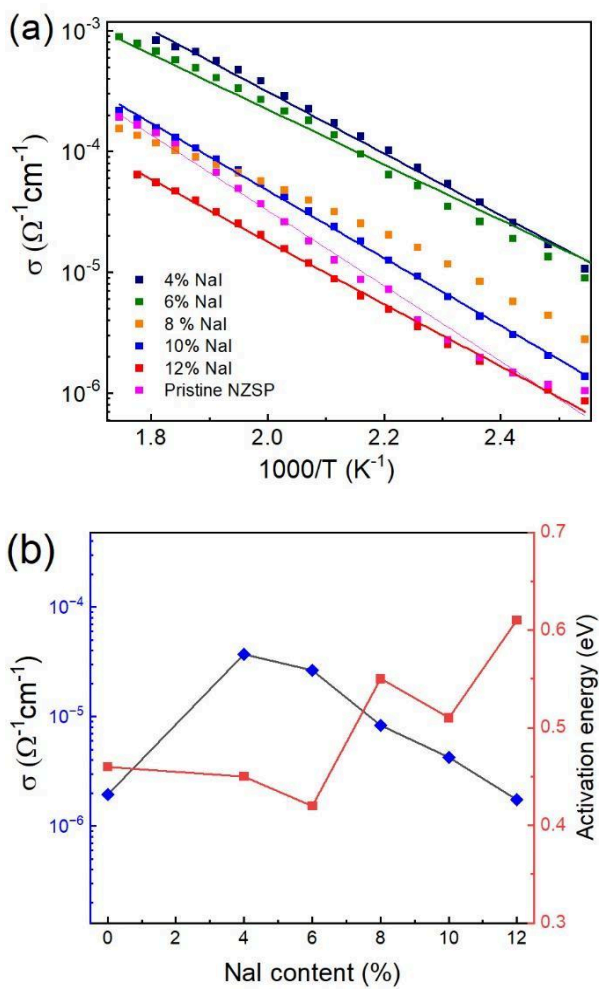

*Fig-4: (a) Temperature-dependent Ionic conductivity of pristine NZSP sample and NZSP-NaI composite (with varying weight percentages of NaI): (b) Ionic conductivities at various weight percentages of NaI in NZSP-NaI composite and activation energy*

**Conclusions:**

The NaI-doped $Na_3Zr_2Si_2PO_{12}$ composites were successfully developed by solid-state reaction route. The introduction of alkali salts NaI into NZSP compounds leads to a significant enhancement in ionic conductivity. The composites with 4% NaI content exhibited the highest ionic conductivity of ~ 3 x $10^{-4}$ $\Omega^{-1}cm^{-1}$ at 200°C which is one order of magnitude rise as compared to pristine NZSP. Such a significant rise in ionic conductivity could lead these composites as suitable candidates for solid-state devices.

**ACKNOWLEDGEMENT**

This work was supported by Shiksha Sopan under the program of student research exposure. We would like to thank Prof. H C Verma for the student research program and his guidance. Also, we would like to thank Prof. Y. N. Mohapatra, Department of Physics, IIT Kanpur, for XRD and FESEM characterization support. Authors Aditya Sekhar and Nikesh Lilani would like to thank Shiksha Sopan for providing the internship.

**Conflict of Interest:** On behalf of all authors, the corresponding author states that there is no conflict of interest.

**Data Availability:** The datasets generated during and/or analysed during the current study are available from the corresponding author on reasonable request.